\documentclass[12pt]{iopart}
\usepackage[round]{natbib}
\usepackage{graphicx}
\usepackage{enumitem}
\begin{document}
%\bibliographystyle{dcu}
%\citationstyle{dcu}

\title[]{The role of vowel and consonant onsets in neural tracking of natural speech}

\author{Mohammad Jalilpour Monesi$^{1,2}$, Jonas Vanthornhout$^2$,  Hugo Van hamme$^1$, Tom Francart$^2$}

\address{$^1$KU Leuven, Dept. of Electrical engineering (ESAT), PSI, Leuven, Belgium\\
  $^2$KU Leuven, Dept. Neurosciences, ExpORL, Leuven, Belgium}
\ead{mohammad.jalilpourmonesi@esat.kuleuven.be, jonas.vanthornhout@kuleuven.be, tom.francart@kuleuven.be, hugo.vanhamme@esat.kuleuven.be}
\vspace{10pt}
% \begin{indented}
% \item[]August 2017
% \end{indented}

% title suggestions:
%Suggestion 1: The Role of Vowel-Consonant Onsets in Neural Tracking
%Suggestion 2: Vowel and consonant onsets are all you need
%Suggestion 3: Vowel-Consonant Onsets: A Key Component in Neural Speech Tracking

\begin{abstract}
To investigate how the auditory system processes natural speech, models have been created to relate the electroencephalography (EEG) signal of a person listening to speech to various representations of the speech. Mainly the speech envelope has been used, but also phonetic representations. We investigated to which degree of granularity phonetic representations can be related to the EEG signal. We used recorded EEG signals from 105 subjects while they listened to fairy tale stories. We utilized speech representations, including onset of any phone, vowel-consonant onsets, broad phonetic class (BPC) onsets, and narrow phonetic class (NPC) onsets, and related them to EEG using forward modeling and match-mismatch tasks. In forward modeling, we used a linear model to predict EEG from speech representations. In the match-mismatch task, we trained a long short term memory (LSTM) based model to determine which of two candidate speech segments matches with a given EEG segment. Our results show that vowel-consonant onsets outperform onsets of any phone in both tasks, which suggests that neural tracking of the vowel vs. consonant exists in the EEG to some degree. We also observed that vowel (syllable nucleus) onsets are better related to EEG compared to syllable onsets. Finally, our findings suggest that neural tracking previously thought to be associated with broad phonetic classes might actually originate from vowel-consonant onsets rather than the differentiation between different phonetic classes.

\end{abstract}

%
% Uncomment for keywords
\vspace{2pc}
\noindent{\it Keywords}: EEG, LSTM, vowel, consonant, match-mismatch, broad phonetic classes
%
% Uncomment for Submitted to journal title message
%\submitto{\JPA}
%
% Uncomment if a separate title page is required
%\maketitle
% 
% For two-column output uncomment the next line and choose [10pt] rather than [12pt] in the \documentclass declaration
%\ioptwocol
%

\section{Introduction}

% Importance of decoding speech from EEG for diagnostic tests, hearing loss

The ability to decode speech from electroencephalography (EEG) signals has the potential to revolutionize diagnostic tests for hearing loss and other speech-related disorders. EEG allows for a non-invasive and cost-effective method of measuring brain activity related to speech perception and production. Despite numerous studies on auditory EEG, our understanding of neural mechanisms of underlying speech perception is still limited, and there is a need for developing new techniques to relate EEG to speech perception or hearing.

% common approaches : linear forward, backward, and hybrid models
A common approach involves recording EEG signals from individuals as they listen to natural running speech, and then attempting to relate the recorded EEG data to the presented speech. There are various methods for relating EEG to speech, but the most common approaches are backward, forward, and hybrid modeling. In backward modeling, usually, a model is used to reconstruct speech representations, such as the speech envelope or spectrogram, from EEG signals \citep[e.g.][]{crosse_multivariate_2016,vanthornhout_speech_2018,verschueren_neural_2019}. In forward modeling, EEG responses are predicted from speech representations \citep[e.g.][]{diliberto_low-frequency_2015,crosse_multivariate_2016,lesenfants_data-driven_2019}. In both approaches, the correlation between the reconstructed or predicted and original signal is used as a proxy for the model's ability to relate EEG to the speech stimulus. In hybrid modeling, such as canonical correlation analysis (CCA), both speech and EEG signals are transformed to a common space where they have the maximum correlation with each other \citep{de_cheveigne_decoding_2018,de_cheveigne_multiway_2019}. Similarly, the correlation between the transformed EEG and the transformed speech is used as an evaluation metric for the hybrid model's ability to relate EEG to speech. It has been shown that these approaches can be used to assess speech understanding from EEG \citep{vanthornhout_speech_2018,lesenfants_predicting_2019,di_liberto_neural_2021}.

% match-mismatch task
\citet{de_cheveigne_decoding_2018} introduced a novel paradigm called the match-mismatch task to relate EEG to speech. The task involves determining whether a segment of EEG matches a segment of speech. In other words, if the EEG segment was recorded while the subject was listening to the speech segment, then the (EEG, speech) pair is considered matched. Conversely, if the EEG was not recorded during the presentation of the speech segment, the (EEG, speech) pair is considered mismatched. Recently, models based on artificial neural networks (ANNs) have shown promising results on this task, outperforming the linear methods by a large margin. \citet{monesi_lstm_2020} introduced a long short term memory (LSTM) based model to determine whether an (EEG, speech) pair is matched or mismatched. Similarly, \citet{accou_modeling_2021} introduced a model based on dilated convolutional layers with similar results as those of the LSTM-based model. In a follow-up study, the authors showed that it is possible to predict speech intelligibility based on match-mismatch classification accuracy of the dilated model \citep{accou_predicting_2021}. In another study, which uses the same match-mismatch task and the same dilated model from \citep{accou_modeling_2021}, \citet{puffay_2022_fo} showed that it is possible to relate the fundamental frequency of the voice (f0) to EEG.

% I (Hugo) added this section, trying to define "phonetic information" and motivate our choices.
Most of the abovementioned studies use the envelope or the spectrogram as the representation for speech. The latter is a very informative representation since automatic speech recognition systems are capable of accurately transcribing from spectral representations. The former is, however, less informative since it merely represents the instantaneous broadband energy. Yet, this representation of speech is quite effective in the match-mismatch task \citep{monesi_lstm_2020}. However, it is unclear what level of detail in the speech representation is sufficient to be successful at the match-mismatch task. 
Rather than modifying the spectral resolution to control the informativeness of the speech representation, we opt for a linguistic representation of speech, which is more interpretable towards the end goal of diagnosing speech and language-related disorders. 
A first choice is that speech will be represented by phonetic classes, ranging from 37 narrow phonetic classes over five broad phonetic classes to the binary vowel versus consonant distinction. Notice that our stimulus data is (automatically) annotated with 37 narrow phonetic classes that collapse allophonic variation in a single class, though pronunciation variants among the 37 narrow classes are automatically identified. Our data hence does not allow us to perform an accurate study about the perception of articulatory features such as aspiration. %manner and place of articulation. 
We restrict phonetic detail to a hierarchy of classes as mentioned above.
 A second choice regards the temporal encoding of phonetic representations. Apart from the speech versus silence property, phonetic representations are encoded at the phone onset. With this choice, repeated occurrences within the same phonetic class are still encoded. By contrast, an encoding of the class identity for the whole phone duration would merge consecutive phones of the same class into a single segment and onset information would be lost. The onset-based encoding hence allows to quantify the impact of class width definition. 
A third choice regards temporal resolution. We compare the high-rate phone-level with the low-rate syllable-level encoding of onsets.
We apply linear models and artificial neural networks to relate the above speech representations to the EEG signal. In the present study, it is unclear whether these models relate language-specific phonemic responses or acoustic phonetic responses from the brain with the discrete speech representations described above. It will also be clear from our analysis below that this distinction is hard to make, given the small differences in match-mismatch task performance observed between fine and broad phonetic classes. For this reason, we will refer to the above representations as \textit{phonetic information}.

There are previous studies relating phonetic information to EEG (or electrocorticography (ECoG)) in which subjects listen to natural running speech. \citet{mesgarani_phonetic_2014} found that different locations in the superior temporal gyrus (STG) had different responses to different broad phonetic classes (referred to as phonetic features in the original study). \citet{khalighinejad_dynamic_2017} found that phones (called phonemes in the original study) of various phonetic categories (such as vowels, plosives, nasals, and fricatives) encode different phonetic distinctions across different time intervals with regard to phone onsets. \citet{di_liberto_low-frequency_2015} reported that adding broad phonetic classes (called phonetic features in the original study) on top of the spectrogram yields better EEG predictions in linear forward modeling. In multiple follow-up studies \citep[e.g.][]{di_liberto_indexing_2017,di_liberto_cortical_2018,di_liberto_atypical_2018,liberto_emergence_2023}, authors reported that broad phonetic classes together with spectrograms yield better EEG predictions (higher correlations) than only using the spectrogram. Similar results were reported by \citet{lesenfants_data-driven_2019}. \citet{prinsloo_general_2022} reported that narrow phonetic classes (phonemes) representations contribute to EEG prediction when speech is intelligible for listeners. In these studies that have investigated neural tracking of phonetic information, simpler speech representations such as phone onsets or vowel vs. consonant onsets have not been utilized. As a result, it is unclear whether the reported increase in EEG prediction accuracy is due to the extra information provided in these phonetic classes or simply the presence of vowel vs. consonant or phone onsets. \citet{daube_simple_2019} reported that adding articulatory features on top of the spectrogram improves EEG predictions of the linear model in forward modeling. Interestingly, they observed a comparable increase in performance when using the phone (phoneme) onsets together with spectrograms. As a result, the authors concluded that the improved prediction accuracy is primarily attributed to the temporal information provided by phone onsets rather than the phonetic information present in different phonetic classes.

While there are other studies that relate EEG to phonetic representations, such as distinguishing between vowels and consonants, these studies have either used unnatural speech stimuli or measurement paradigms that involve articulating or imagining speech rather than simply listening to speech stimuli. For example,  \citet{parhi_classifying_2021} developed two models based on CNN and LSTM neural networks to classify imagined vowels (i, a, u) from EEG. They showed that using only frontal lobe EEG channels, classification accuracy exceeded 85\% for all of the subjects. In another study, \citet{banerjee_significance_2022} showed that applying principal component analysis (PCA) to a CNN-based feature vector increases the classification of imagined vowels. In \citep{mahapatra_multiclass_2022}, authors combined temporal and spatial features using CNNs in a deep learning model to classify imagined speech to five vowels and six words. They obtained an average accuracy of 96.49\%, which is 11\% higher than the state of the art model \citep{sarmiento_recognition_2021}. In a different paradigm where subjects listened to separate vowels or syllables in the form of consonant-vowel (CV), \citet{wang_using_2012} classified 8 consonants and 4 vowels from the recorded EEG. More specifically, they showed that it is possible to classify distinctive features such as voicing, continuant, place, height, and backness from the recorded EEG. In a similar study, \citet{kovacs_eeg_2017} studied the event-related potentials (ERPs) in response to different consonant classes. They reported that different ERP responses were elicited in response to different phonetic categories (fricatives, plosives, liquids, nasals, and affricates).

% layout of the next sections

In this study, we aim to understand to which degree of granularity phonetic information can be related to the EEG signals recorded from subjects while they listened to a natural running speech. To this end, we tried to relate EEG to speech representations such as the onset of any phone, of vowels, of consonants, of broad phonetic classes (BPC) and of narrow phonetic classes, using two different tasks. In the forward modeling task, we used a linear model to predict EEG from these speech representations, while in the match-mismatch task, we trained an LSTM-based model to classify matched and mismatched (EEG, speech) pairs.

\section{Methods}
\label{sec:methods}

In this section, we will first provide details about the EEG dataset used in this study. Then, we will explain how we processed the EEG data and give an explanation of the speech representations that we extracted from the speech stimulus. Next, we will define the two tasks that were used as an evaluation metric to measure how well the speech stimulus can be related to EEG. The first task is the common approach of predicting EEG from the stimulus, also known as forward modeling. The second task is a recent approach, inspired by the auditory attention decoding literature, to relate EEG to the stimulus using a classification task called match-mismatch. Lastly, we will introduce the LSTM-based model \citep{monesi_lstm_2020,monesi21_interspeech} that was used in the match-mismatch task.

\subsection{Dataset}
EEG signals were recorded from subjects while they listened to a natural running speech in the form of a fairy tale story. Subjects were normal-hearing Flemish speaking in the [18-30] age group. Subjects gave their informed consent, which was reviewed and approved by the Medical Ethics Committee at KU Leuven in Belgium, under reference number S57102. First, participants were screened for a normal hearing with pure tone audiometry and Flemish matrix-test \citep{luts_development_2014}. Only participants with normal hearing (hearing threshold \textless~25~dBHL) went through EEG measurements. 

For this study, we used EEG recordings of 105 subjects who listened to fairy tale stories. For most subjects, we had 7 or 8 recordings. For 9 subjects, we had less than 7 recordings due to measurement errors. Each story was approximately 14 minutes and 30 seconds in length, and the order of presentation was randomized. Participants were given breaks between recordings. The stimuli were presented using the APEX 4 software developed at ExpORL \citep{francart_apex_2008} and delivered binaurally through Etymotic ER-3A insert phones at 62 dBA. To encourage attentive listening, participants were informed that they would be asked questions about each story after completion. EEG signals were recorded using the 64-channel Active-two Biosemi system with a 8192~Hz sampling rate, within an electromagnetically shielded and soundproofed cabin. The EEG recordings were synchronized with the stimuli after each session. For more detailed information about the dataset, please refer to \citep{K3VSND_2023}.

\subsection{Preprocessing}
\textbf{EEG:} The EEG signals were first high-pass filtered at 0.5~Hz for drift correction using a fourth-order Butterworth filter. Matlab's filtfilt function was used to have zero-phase filtering. The recordings were then downsampled from 8192~Hz to 1024~Hz using Matlab's resample function, which includes an anti-aliasing low-pass filter prior to resampling. After removing artifacts from the EEG signals using the multi-channel Wiener filter \citep{somers_generic_2018}, all EEG channels were re-referenced to the common average of the channels. The EEG signals were then downsampled again to 64~Hz using Matlab's resample function. The resulting EEG signals had a bandwidth between 0.5 and 32~Hz.

\begin{figure}[htp]
	
	\begin{minipage}[b]{1\linewidth}
		\centering
		\centerline{\includegraphics[width=\textwidth]{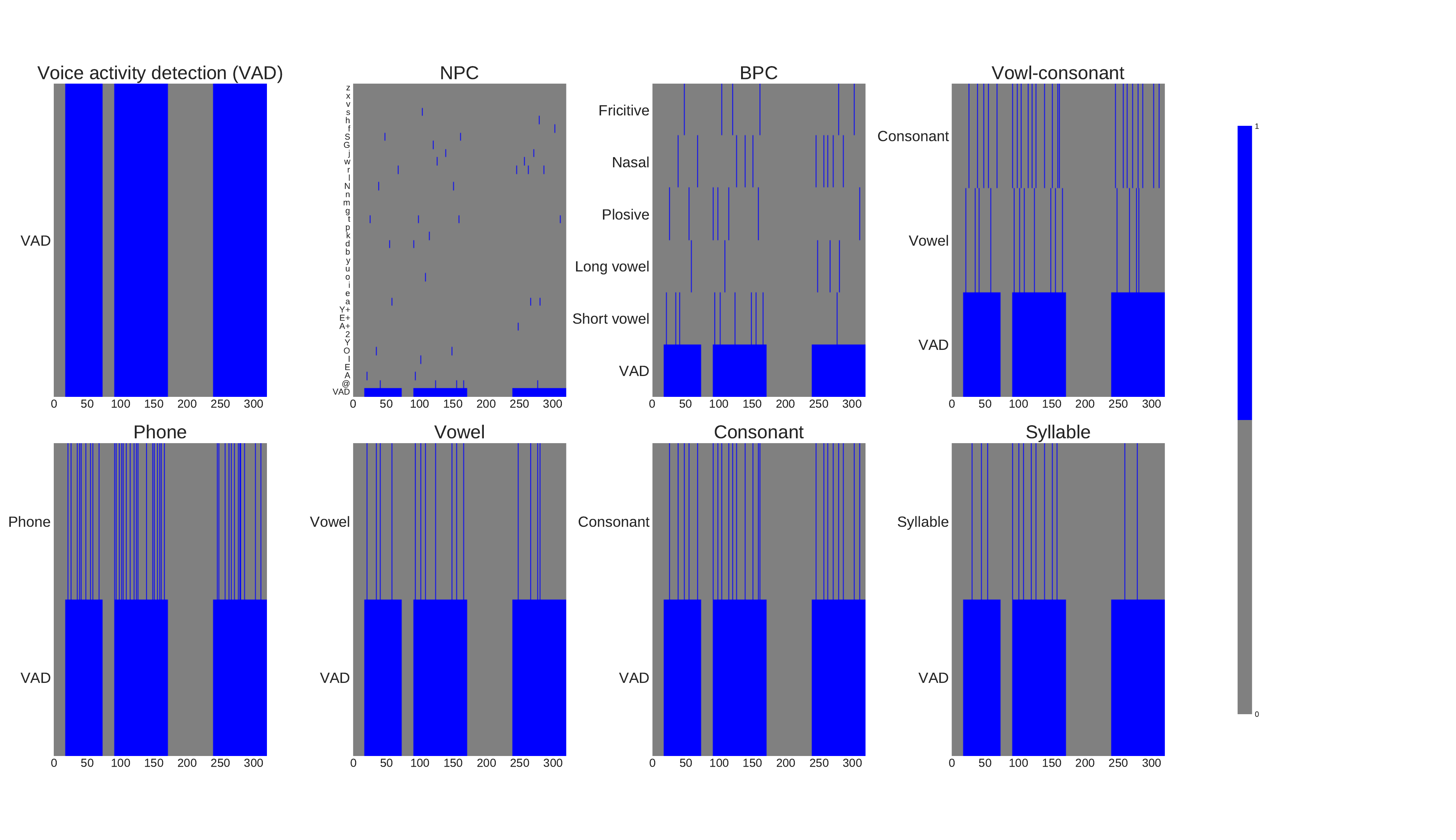}}
		%  \vspace{2.0cm}
		%	\centerline{(a) Result 1}\medskip
	\end{minipage}

	\caption{The speech representations employed in this study are displayed over a five-second interval. Note that VAD is added to each of the other representations as baseline speech information.  With the exception of the syllable onsets, all of the speech representations  were calculated based on narrow phonetic classes (NPC) onsets.}
	\label{fig:speech_features}
\end{figure}

\textbf{Speech representations:} A summary and visualization of each speech representation is provided in table \ref{spch_table} and figure \ref{fig:speech_features}, respectively. The following speech representations were used in this study.

% \begin{description}
% \item[1.xyz] \hfill \\ this is 
% \item[2. uyx] \hfill \\ that is
% \end{description}

% VAD

\begin{enumerate}
    \item[1.]  Voice activity detection (VAD): This one-dimensional feature distinguishes speech from silence, with a value of 1 indicating speech and 0 indicating silence. This distinction was shown by \citet{monesi21_interspeech} to be the main contributor to match-mismatch accuracy. As a result, we used this feature as baseline information and thus added it as a first dimension to all of the following speech representations. It is worth noting that in this study, VAD was calculated based on phone alignment rather than speech energy thresholding.

% NPC
\item[2.] Narrow phonetic classes (NPC) onsets: Using forced alignment developed especially for reading tasks \citep{duchateau_2009}, we segmented each story into a sequence of phone units. The story text is mapped to a hidden Markov model that allows for disfluencies by the narrator, including word restarts and repetition of and skipping over several words. The HMM also chooses the most likely among several pronunciations. The pronunciation model accounts for coarticulation effects such as degemination. The model, however, does not distinguish in fine allophonic detail (e.g. realization of /r/ as uvular trill, alveolar trill or alveolar flap) and hence we term the units as \textit{narrow phonetic classes}. % based on the International Phonetic Alphabet (IPA). 
Then for each of the 37 NPC, we converted the symbol to a 37-dimensional one-hot vector for each time sample. More specifically, the encoding was done at phone onsets to preserve the duration between phones. The following speech representations (except for the syllable onsets) are all derived from NPC onsets by decreasing the granularity level of phonetic information.

% BPC
\item[3.] Broad phonetic classes (BPC) onsets: Following \citet{lesenfants_data-driven_2019}, we started from NPC and grouped vowels into 2 categories (short vowels and long vowels) and consonants into 3 categories (plosives, fricatives, and nasals/approximants) based on the manner of articulation. As a result, BPC is a 5-dimensional one-hot vector for each time sample. Note that compared to NPC, BPC contains less phonetic information.

% vowel/consonant
\item[4.] Vowel-consonant onsets: We created a two-dimensional one-hot vector from BPC onsets by grouping vowels and consonants. Note that this representation is derived from BPC onsets and contains less phonetic information compared to BPC onsets.

\item[5.] Phone onsets: we combined vowels and consonants in the vowel-consonant onsets representation to one category called phone onsets. This one-dimensional representation does not contain the discrimination between vowels and consonants.

% vowel
\item[6.] Vowel onsets: This representation is directly derived from vowel-consonant onsets by excluding consonant onsets. As a result, this one-dimensional feature contains only vowel (syllable nucleus) onsets.

% consonant
\item[7.] Consonant onsets: This representation is directly derived from vowel-consonant onsets by excluding vowel onsets. As a result, this one-dimensional feature contains only consonant onsets.

% syllable
\item[8.] Syllable onsets: This representation is one at syllable onsets and zero otherwise.
\end{enumerate}

The objective behind developing simplified phonetic representations is to examine the level of granularity at which the model can effectively leverage phonetic information to relate EEG to speech. All the speech representations were calculated at a 64~Hz sampling rate so they are synchronous with the EEG.

% tabel 
\begin{table}[]
	\caption{\label{spch_table}Summary of speech representations used in this study.}
	\begin{indented}
		\item[]\begin{tabular}{@{}llll}
			\br
			Speech representation&Dimension&Explanation\\
			\mr
			VAD& 1      & One for speech activity, zero otherwise\\
			NPC& 37 & Narrow phonetic classes onsets \\
			BPC& 5        & Contains onsets of five categories: short vowels, long vowels, \\
			   &	&	plosives, fricatives, nasals and approximants \\
			Vowel-consonant &2  & Vowel and consonant onsets\\
			Phone &1            &Phone onsets, it is one for the onsets of any phone \\
			& &	and zero otherwise\\
			Vowel& 1      & Vowel onsets \\
			Consonant& 1        &Consonant onsets\\
			Syllable& 1       &Syllable onsets\\
			\br
		\end{tabular}
	\end{indented}
\end{table}

% figure

% split and normalization

\textbf{Split and normalization:} Each individual EEG recording (for most subjects, we had 7 or 8 recordings of approximately 15 minutes each) was split into a training set (80\%), a validation set (10\%), and a test set (10\%). More specifically, the data was split into (40\%, 10\%, 10\%, 40\%) portions. The train set contained 40\% of the recording at the start and 40\% of the data at the end of the recording. Validation and test sets were selected from the middle of the recording in order to prevent any artifacts that might exist at the beginning or end of the recording. Lastly, each recording was mean-variance normalized. The mean and variance of the training set were used to normalize training, validation, and test sets.

\subsection{Linear forward modeling}

We used forward modeling \citep[e.g.][]{crosse_multivariate_2016,lesenfants_data-driven_2019} as one of the evaluation methods to relate EEG to speech. In this approach, a linear model was used to predict the recorded EEG from the speech stimulus. After predicting the EEG, the correlation between the recorded EEG and the predicted EEG was used as a measure of performance. We used ridge regression as a regularization method. The analytical solution of the linear forward model with ridge regression regularization is as follows:

\begin{equation}\label{eqn:Rigde regression}
W = (S^{T}S + \lambda I)^{-1}S^{T}R
\end{equation}
$S$ is the lagged time series of the speech representation and hence takes the delay of the brain responses with regard to the stimulus into account. The integration window was set to 400 milliseconds, meaning that speech samples up to 400 milliseconds prior to the EEG time sample were used for prediction. $R$ represents the recorded EEG signal and $I$ is the identity matrix . Finally, $W$ is the linear model, which is also known as the temporal response function (TRF). The value of $\lambda$ is usually determined based on a validation set or k-fold cross-validation. In this study, the value of the $\lambda$ was chosen based on the performance of the linear model on the validation set. The Spearman correlation between the predicted EEG channel and the actual EEG channel was used as a performance metric for this task.

\subsection{Deep-learning model in a match-mismatch task}

\begin{figure}[htb]
	
	\begin{minipage}[b]{1\linewidth}
		\centering
		\centerline{\includegraphics[width=\textwidth]{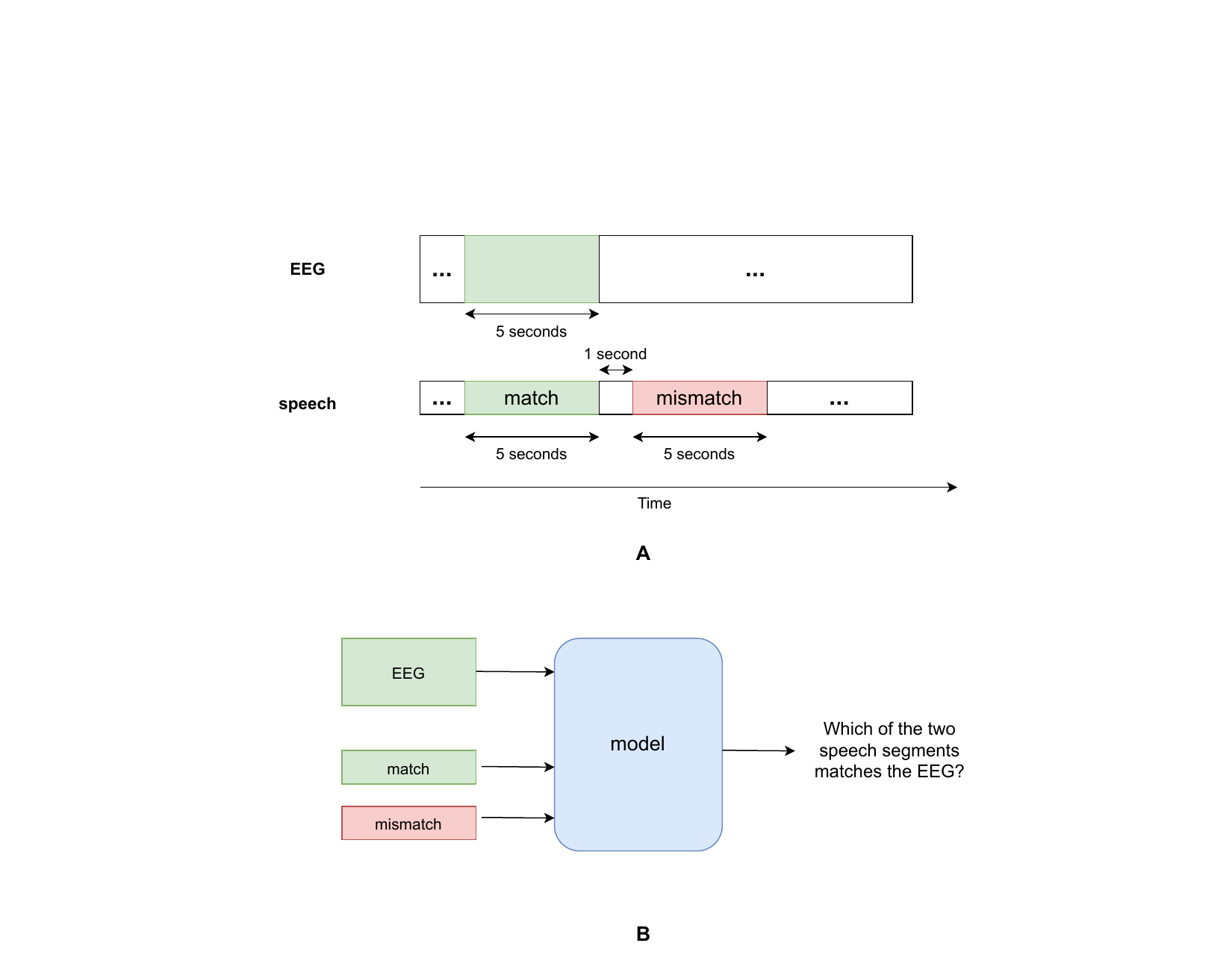}}
		%  \vspace{2.0cm}
		%	\centerline{(a) Result 1}\medskip
	\end{minipage}

	\caption{A) Extracting matched and mismatched speech segments with respect to an EEG segment of 5 seconds (decision window). B) match-mismatch classification task where a model is trained to determine which of the two given speech candidates matches the provided EEG.}
	\label{fig:match_mismatch}
\end{figure}

% add (train, validaiton, test) split info
In this section, we will explain the second task used in this study to relate EEG to the speech stimulus. This binary classification task, referred to as the match-mismatch task, has been used in recent studies to relate EEG to the presented speech stimulus \citep[e.g.][]{cheveigne_auditory_2020,monesi_lstm_2020,accou_predicting_2021}. 
The task involves determining whether a segment of a recorded EEG is aligned with a presented speech stimulus.
The EEG and speech segments are considered a match if the EEG was recorded while the speech was being presented, meaning they are temporally aligned. 
Conversely, the EEG and speech segments are considered a mismatch if there is no temporal alignment between the two. 
The procedure for extracting matched and mismatched (EEG, speech) pairs is illustrated in Figure \ref{fig:match_mismatch}.A. It depicts (EEG, speech) pairs of 5-second length (referred to as the decision window), extracted with 80\% overlap. For the mismatched (EEG, speech) pair, the speech is taken one second after the end of the EEG segment (in the future).

After defining matched and mismatched pairs, we can now define the final match-mismatch classification task. As illustrated in figure \ref{fig:match_mismatch}.B, the task is a binary classification where the model is trained to determine which of two candidate speech inputs matches the input EEG segment. 
It should be noted that always one of the speech inputs is a matched pair with the EEG input, and the other one is mismatched. Moreover, the position of the matched and mismatched speech inputs is alternated to prevent the model from consistently identifying one of the inputs as matched.

In order to train a generalizable model, the mismatched speech (similar to hard negative sampling) should be difficult enough. Therefore, the mismatched speech segment is taken from the same stimulus and also temporally close to the matched speech. In this way, the mismatched speech has some general similarities to the matched speech, such as originating from the same stimulus speaker or similar listener fatigue. Thus, the model needs to rely on the content of the two candidate speech segments to determine which one is matched.

\begin{figure*}[htb!]
	
	%\begin{minipage}[b]{1\linewidth}
	\centering
	%\centerline{\includegraphics[width=\textwidth, height=4cm]{lstm_jne.png}}
	%\centerline{\includegraphics[width=\textwidth, height=4cm]{lstm_crop.pdf}}
        \centerline{\includegraphics[width=\textwidth, height=4cm]{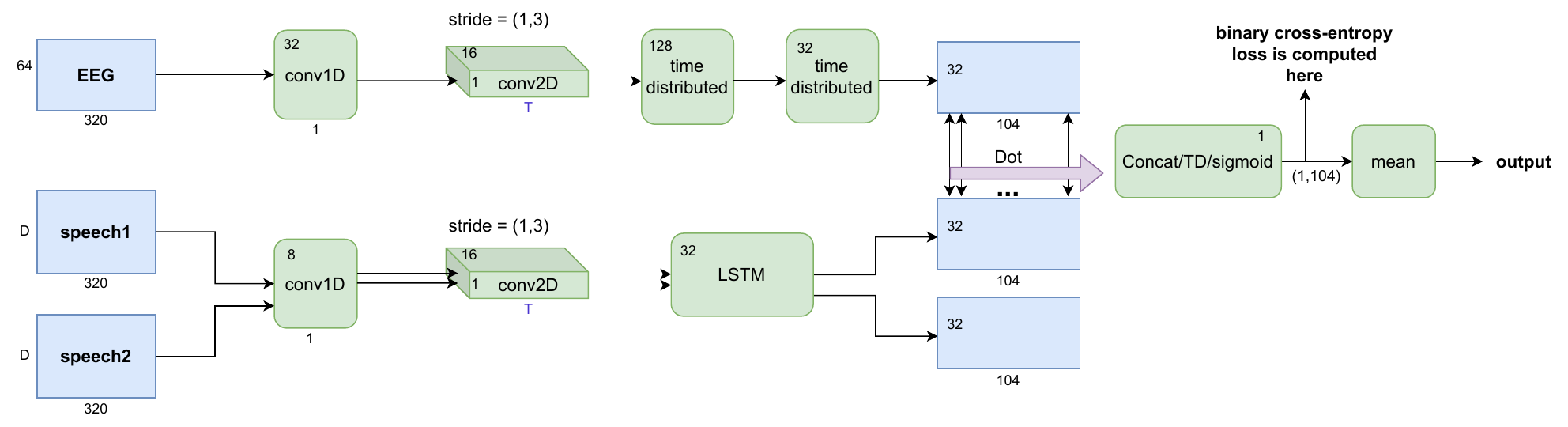}}
	%  \vspace{2.0cm}
	%	\centerline{(a) Result 1}\medskip
	%\end{minipage}
	
	%
	\caption{The LSTM-based model \citep{monesi21_interspeech} that is used in match-mismatch classification. The \textit{Dot} layer applies a normalized dot product (cosine similarity) between EEG and speech representations for each time step. The green color indicates a model layer, while the blue color indicates the input or output of a layer. The selected hyperparameters, such as kernel size or number of units, are shown for each layer. The number of frames at the output( number 104 in the figure) corresponds to a conv2D kernel size of 9 (T=9).}
	\label{fig:LSTM model}
\end{figure*}

\textbf{Model}: We have used an LSTM-based model introduced in \citep{monesi_lstm_2020} and slightly adjusted in \citep{monesi21_interspeech} to do the match-mismatch task. The architecture of this model is shown in figure \ref{fig:LSTM model}. 
The model includes two networks to project each of the inputs into an embedding space. The idea is that in the embedding space, EEG and matched speech have similar representations for each time frame, while EEG and mismatched speech have dissimilar representations in the embedding space. %By similarity we mean how much the two embedded vectors are aligned using the cosine similarity measure. 
Here, the similarity is measured as cosine similarity, which is calculated for each time frame with the assumption that there is enough information to align these short segments of EEG and speech. 
We opt for an LSTM in the speech-related network path because apart from being able to model non-linearities, it can also model the delayed brain response observed in the EEG signals. Ideally, the LSTM layer has the capability to introduce a delay to the speech input and therefore synchronize it with the EEG stream \citep{monesi_lstm_2020}. The layers in the speech path are shared between the two speech candidates.

The hyperparameters of the model are shown in figure \ref{fig:LSTM model}. These hyperparameters were tuned for mel spectrogram representation in a previous study \cite{monesi21_interspeech}. In this study, only the kernel size of the conv2D layer was tuned to have an appropriate receptive field for each speech representation. The model has about 94000 learnable parameters in total. During the training, we used 30 epochs with early stopping. The code of the model is provided at https://github.com/exporl/match-mismatch-LSTM.

% LSTM figure

\section{Results}
\label{sec:results}

We aim to identify which phonetic information in the presented speech stimulus can be related to the EEG response. We have defined eight speech representations with varying levels of phonetic information, as described in the methods section. 
In the first part of this section, we will show the results of relating these speech representations to the recorded EEG using linear forward modeling. To do so, correlation scores between predicted EEG channels and actual EEG channels will be reported. Temporal response functions (TRFs) as well as correlation and TRF topoplots will be shown for further interpretation. In the second part of this section, the results of relating EEG to these speech representations in the match-mismatch task will be discussed. In order to determine whether there is a significant difference between the means of two groups, we used a signed-rank Wilcoxon test throughout this section. All the p-values are corrected for multiple comparisons using the Holm-Bonferroni method \citep{holm_simple_1979}.

\subsection{Relating phonetic information to EEG using linear forward modeling}

With the linear forward modeling approach, EEG signals are predicted from speech representations (regression task) using a linear model. Then, the correlation between the actual EEG and the predicted EEG is used as a proxy for neural tracking of speech. Similar to \citet{lesenfants_data-driven_2019}, we included 27 fronto-temporal channels. 

The phonetic information within each speech representation was systematically controlled, varying from the lowest level of granularity (phone onsets) to the highest level (NPC onsets). In terms of phonetic information, we have the relation: phone $<$ vowel-consonant $<$ BPC $<$ NPC. 
Also, each speech representation contains the previous representation's information. For example, BPC contains all the information available in the vowel-consonant representation plus some extra phonetic information, related to manner of articulation for consonants and duration for vowels. %, incrementally increasing the amount of information with each representation. 

% show corr and explanations

\begin{figure}[htb]
	
	\begin{minipage}[b]{1\linewidth}
		\centering
		\centerline{\includegraphics[width=\textwidth]{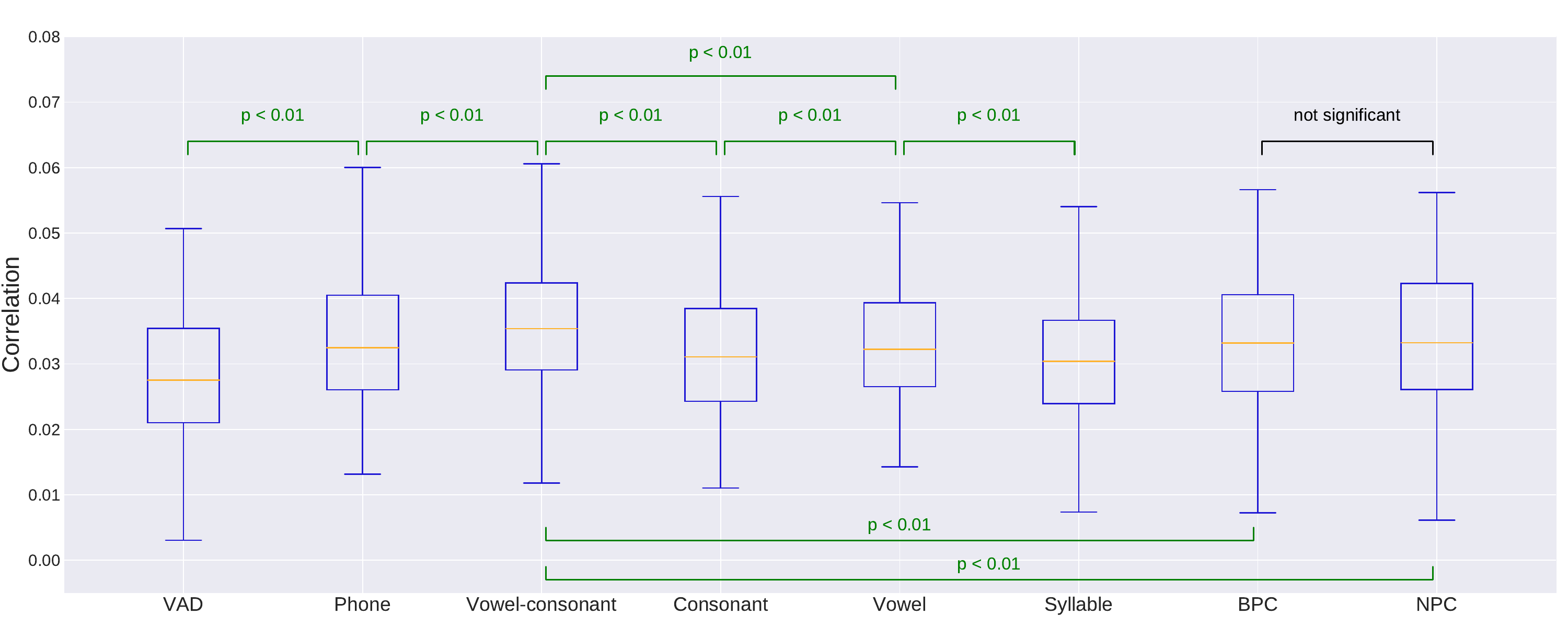}}
		%  \vspace{2.0cm}
		%	\centerline{(a) Result 1}\medskip
	\end{minipage}

	\caption{The Spearman correlations of the linear model for phonetic representations. Box plots are shown over 105 subjects. The model is trained separately for each subject (subject-specific model). Statistical significance tests are only shown for important pairs to simplify the figure. P values are corrected for multiple comparisons using the Holm-Bonferroni method.}
	\label{fig:corr}
\end{figure}

Correlation scores between predicted EEG and actual recorded EEG signals are shown in figure \ref{fig:corr} for each speech representation. 
We see that all the speech representations outperform the baseline VAD representation (all p $<$ 0.01). This shows that the model is able to use the extra phonetic information on top of speech vs. silence to make better EEG predictions. 
Interestingly, vowel-consonant onsets performs better than phone onsets (p $<$ 0.01). This means that providing vowel and consonant discrimination helps the model to have a better EEG prediction. As shown in figure \ref{fig:corr}, vowel-consonant onsets outperforms vowel onsets and consonant onsets (p $<$ 0.01). This suggests that both vowel and consonant onsets contribute to EEG predictions. Also, notice that vowel onsets are better related to EEG than consonant onsets (p $<$ 0.01).

We also included syllable onsets in our experiments to compare it with vowel (syllable nucleus) onsets. As seen in the figure, vowel onsets outperform syllable onsets (p-value $<$ 0.01). Finally, we observe that adding more phonetic information on top of vowel-consonant onsets does not result in an improvement in EEG predictions. BPC and NPC representations perform worse (p $<$ 0.01) than vowel-consonant onsets despite having more phonetic information. In theory, the linear models using the BPC and NPC representations should be able to achieve a fit to the data that is as good as, if not better than, linear models using the vowel-consonant representations. However, this is not the case due to three causes: (i) We work with finite data sets, which causes variance in estimation and evaluation. The filters in the NPC and BPC models are estimated on fewer data points than in the consonant-vowel model, leading to higher variance. (ii) Regularization - required due to finite data sets - results in models that do not minimize MSE (iii) The models are trained for minimal MSE, but evaluated on correlation.

%In theory, for the BPC and NPC representations, the linear model without regularization should be able to converge to the same solution as that of the vowel-consonant onsets.
%causes model   , the linear model without regularization should be able to converge to the same solution as that of the vowel-consonant onsets. However, the inclusion of regularization, which penalizes model weights, can cause the model to find an alternative set of weights that minimizes the mean squared error (MSE). Also, note that our metric here is the correlation between the original and predicted signal and not the MSE. Additionally, it is worth mentioning that the correlation and cross-correlation matrices are estimated based on fewer data points, leading to a higher variance in the model. Consequently, we do not achieve convergence to the solution observed in vowel-consonant onsets.

%In summary, the lower performance of BPC and NPC representations can be attributed to the influence of regularization, the difference in evaluation metrics, and the smaller sample size used for estimating correlation and cross-correlation matrices.

To conclude, we investigated which phonetic information can be used by the linear model to predict the EEG. The results indicate that all speech representations outperform the baseline VAD representation, with vowel-consonant onsets representation performing the best. The main finding was that vowel-consonant onsets contribute to the predictions of the EEG on top of phone onsets and suggest that neural markers of vowel vs. consonant exist in the EEG signals. The study also finds that vowel onsets are better related to EEG than syllable onsets. Furthermore, increasing the granularity level of phonetic information beyond vowel-consonant does not improve EEG predictions.

% show TRF
\begin{figure}[htb]
	
	\begin{minipage}[b]{1\linewidth}
		\centering
		\centerline{\includegraphics[width=\textwidth]{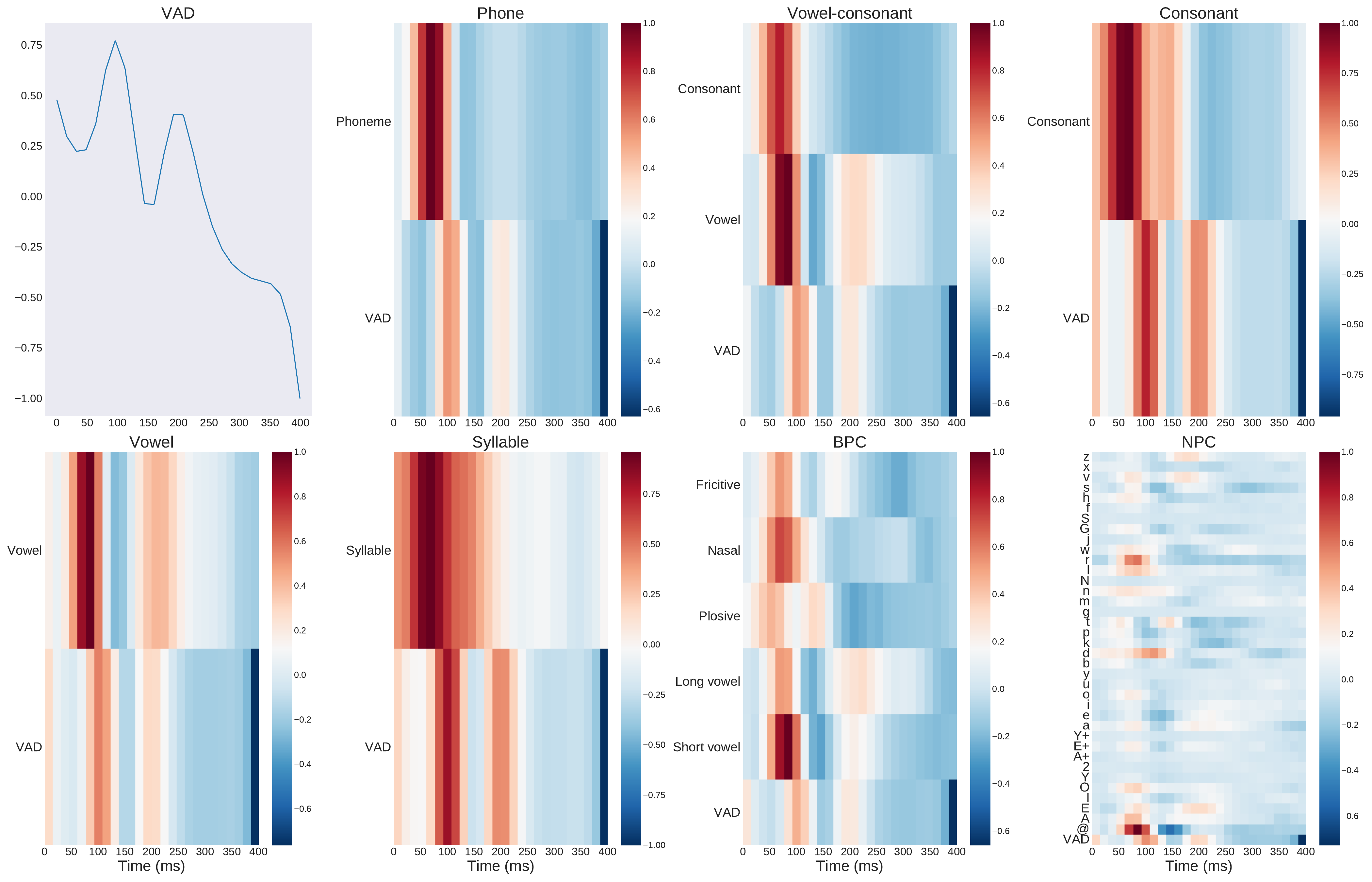}}
		%  \vspace{2.0cm}
		%	\centerline{(a) Result 1}\medskip
	\end{minipage}

	\caption{Average TRFs over 105 subjects are displayed for different speech representations.}
	\label{fig:TRF}
\end{figure}

\subsubsection{Temporal response functions}
Temporal response functions (TRFs) are often used in forward models to analyze the latency of brain responses to speech. In figure \ref{fig:TRF}, average TRFs over all 105 subjects are shown for each of the speech representations. We see that for the baseline VAD representation, there are two positive peaks at 100 ms and 200 ms (VAD subplot in figure \ref{fig:TRF}).

The primary positive peak of the VAD representation occurs at approximately 100 ms, while it occurs earlier for other representations, such as consonants and vowels. For instance, the main and only positive peak for phone onsets occurs around 75 ms (phone subplot in figure \ref{fig:TRF}). Additionally, we observe that the vowel onsets representation peaks later than the consonant onsets representation (vowel-consonant subplot in figure \ref{fig:TRF}).
Unlike vowel onsets, consonant onsets exhibit only one positive peak in the TRF. Furthermore, the first positive peak of the vowel TRF appears to be stronger than that of the consonant in the TRF of vowel-consonant onsets. This indicates that the linear model assigns greater importance to vowel onsets than to consonant onsets. This observation might provide additional evidence to support the notion that EEG tracking of vowel onsets is easier than that of consonant onsets, as demonstrated not only by correlations (as shown in figure \ref{fig:corr}) but also by the weights assigned by the model (vowel-consonant subplot in figure \ref{fig:TRF}).

The TRF for syllable onsets is comparable to that of consonant onsets, which is not surprising considering that most (around 60\%) syllable onsets are consonant onsets rather than vowel onsets. The positive peak in the TRF of both consonant and syllable onsets appears to be broad. Among the BPC categories, short vowels and nasals seem to elicit the strongest TRF responses. Overall, the TRF peaks of BPC are in line with those reported in a similar study by \citet{lesenfants_data-driven_2019}. It is quite difficult to interpret the TRF of a sparse representation such as NPC since some of the classes (phones) occur rarely in the training set. Nevertheless, it seems that vowels yield two positive peaks, and most consonants yield one positive peak, which is in line with the TRF of other representations in this study.

% energy distribution of vowels vs consonants

\begin{figure}[htb]
	
	\begin{minipage}[b]{1\linewidth}
		\centering
		\centerline{(a) \includegraphics[width=\textwidth]{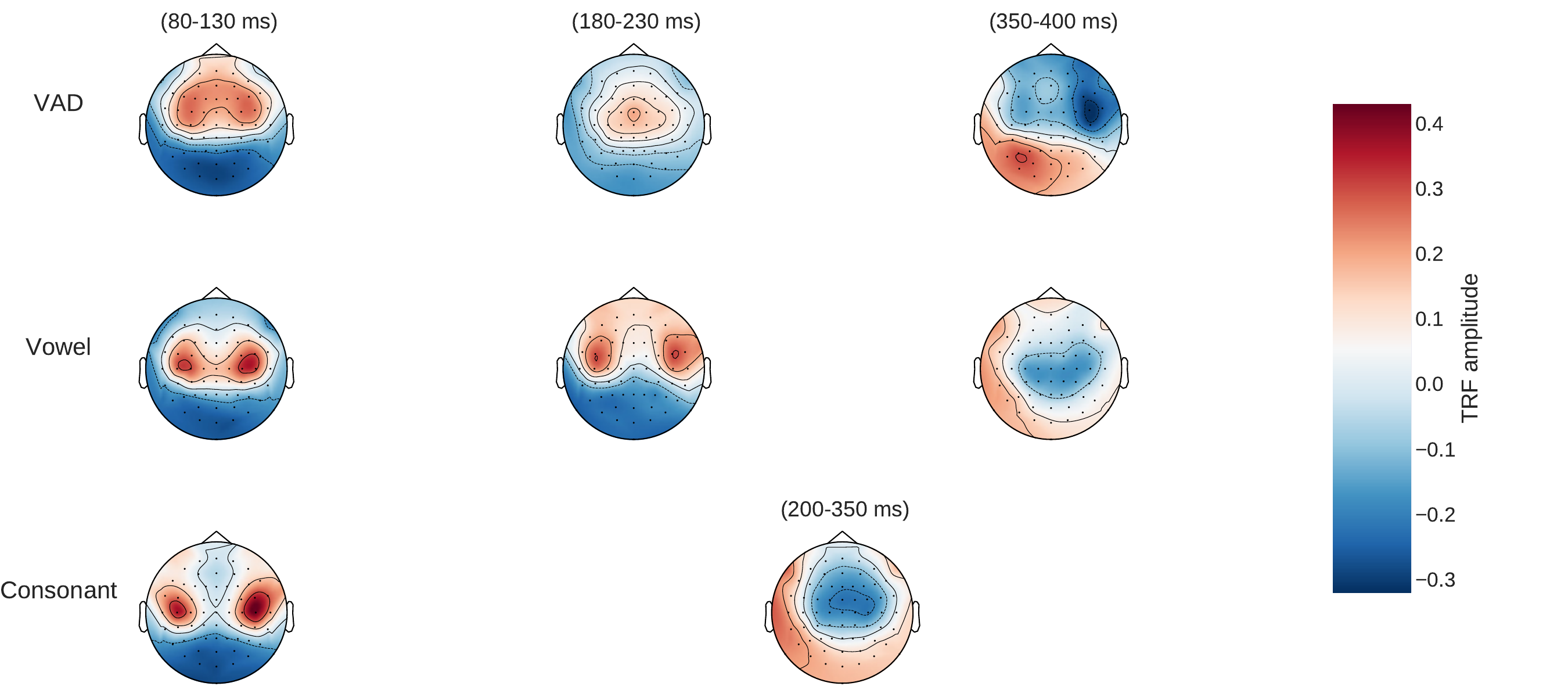}}
            \vspace{1.5cm}
            \centerline{(b) \includegraphics[width=\textwidth]{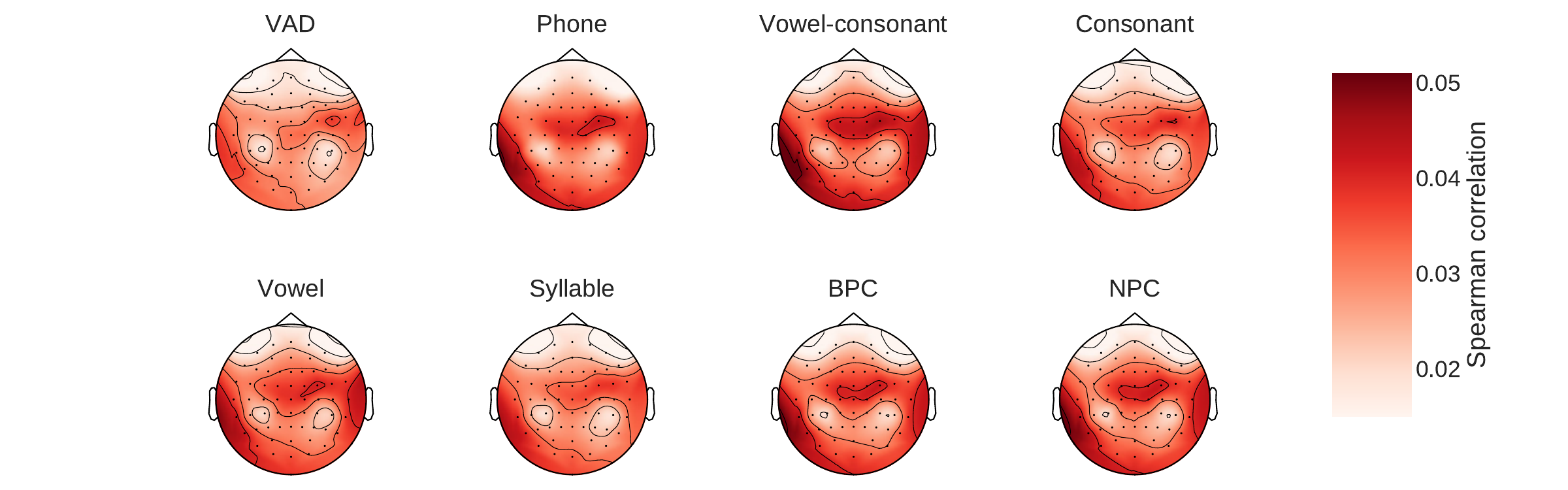}}\medskip
		\vspace{1.0cm}
		%	\centerline{(a) Result 1}\medskip
	\end{minipage}

	\caption{(a) TRF topoplots over 64 channels are shown for VAD, vowel onsets, and consonant onsets (vowel-consonant onset representation is used here). For VAD and vowel onsets the average TRF values are presented over two positive peaks, [80-130 ms] and [180-230 ms], and a negative peak [350-400 ms]. For consonant onsets, TRF values are shown for one positive peak [80-130 ms] and one negative peak [200-350 ms]. (b) Topoplot representation of EEG channel prediction.}
	\label{fig:topo_trf}
\end{figure}

\subsubsection{EEG topoplots}

In order to know which EEG channels have stronger TRF amplitudes during positive and negative peaks, usually TRF topoplots are used. In figure \ref{fig:topo_trf}, TRF topoplots are shown for sub-components of the vowel-consonant representation, namely VAD, vowel onsets, and consonant onsets. VAD and vowel onset TRFs have two positive and one negative peak, while consonant onsets have one positive and one negative peak. As a result, time intervals [80-130 ms], [180-230 ms], and [350-400 ms] are chosen for VAD and vowel onsets, while [80-130 ms] and [200-350 ms] are chosen for consonant onsets.

During the first positive peak interval of [80-130 ms], high positive weights are seen in both temporal and central channels and high negative weights are seen in occipital channels for all three representations.
For the [180-230 ms] time interval, higher model weights are concentrated in central channels for VAD and in temporal channels for vowel onsets. The TRF weight pattern observed during the negative peak interval of vowel onsets ([350-400]) is found to be the same as that observed during consonant onsets ([200-350]). Specifically, high negative weights are noted in the central channels, while high positive weights are noted in the left temporal and occipital channels. For the VAD TRF, the right temporal and frontal channels have negative values, while the left temporal and occipital channels have positive values.

Finally, we also provide EEG prediction topoplots in figure \ref{fig:topo_trf} to see which EEG channels are predicted more correctly than others. Figure \ref{fig:topo_trf}.(b) presents the correlation values per channel, which have been averaged over all subjects. The highest correlations are found in the left and right temporal channels, as well as in some central channels, while occipital and frontal channels have lower correlations. This indicates that the temporal channels are easier to predict, which is plausible as the primary auditory cortex is located in the temporal lobe of the brain.

\subsection{Relating phonetic information to EEG using a deep learning model in the match-mismatch task}
\label{sec:ph}

% preface, general introduction

In this section, we relate the recorded EEG signals to the presented speech in the match-mismatch task using the LSTM-based model shown in figure \ref{fig:LSTM model}. 
First, we trained a subject-independent (SI) model using data from all the subjects. 
Then, we fine-tuned the SI model for each subject. It has been shown in \citep{monesi_lstm_2020} that further fine-tuning on each subject improves the model's performance. 
We trained the model using the gradient descent method with the ADAM optimizer. The model was trained for 30 epochs with early stopping. If the validation loss increased for 5 consecutive epochs, the training would stop. The speech representations used in this section were identical to those utilized in linear forward modeling.

% showing the results and discussing it

\begin{figure}[htb]
	
	\begin{minipage}[b]{1\linewidth}
		\centering
		\centerline{\includegraphics[width=\textwidth]{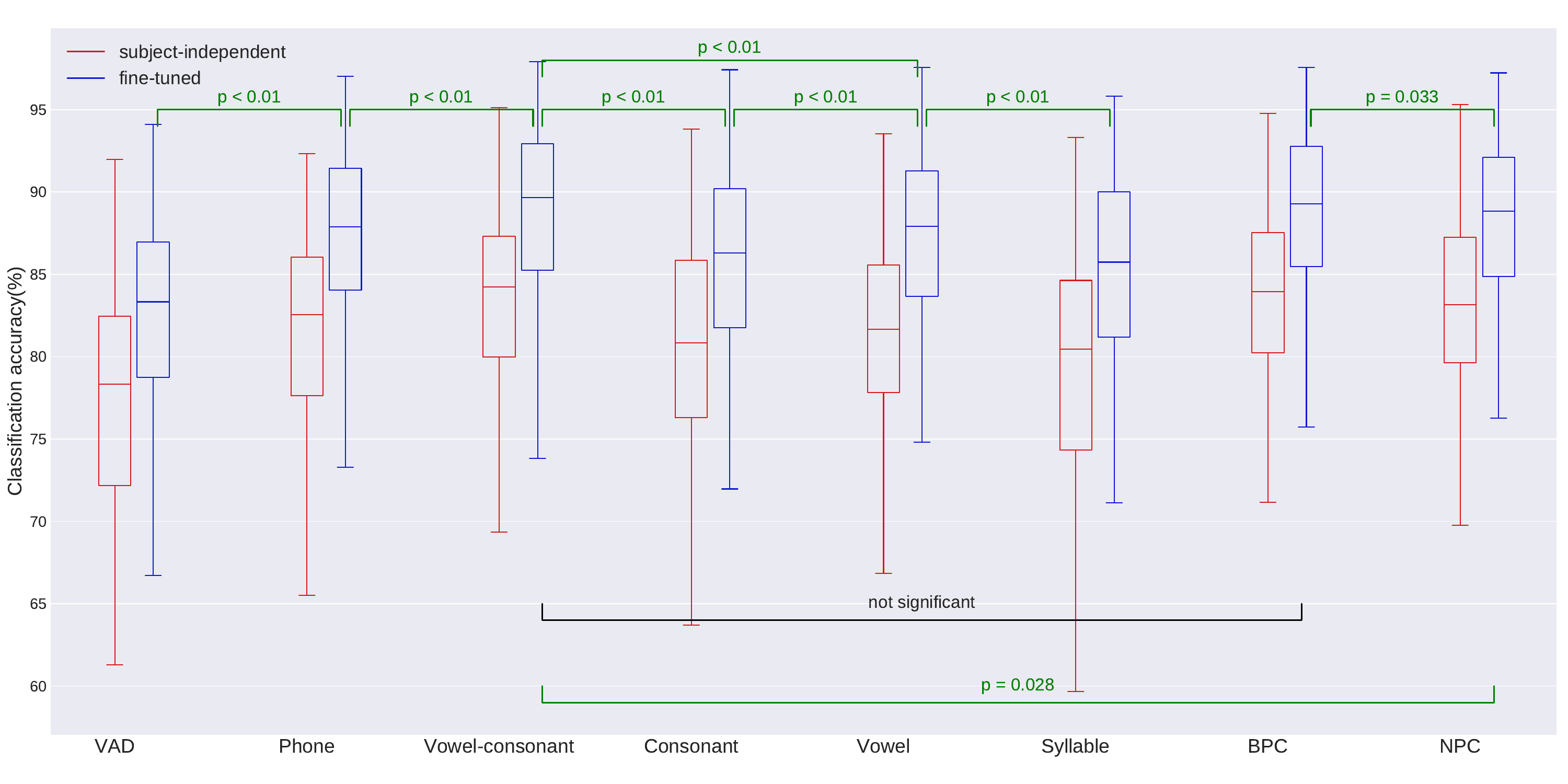}}
		%  \vspace{2.0cm}
		%	\centerline{(a) Result 1}\medskip
	\end{minipage}

	\caption{Match-mismatch accuracy of the subject independent and fine-tuned models for each of the speech representations using decision windows of 5 seconds. Box plots are shown over 105 subjects. Statistical significance tests are only presented between some of the pairs (important pairs such as phone onsets vs vowel-consonant onsets) to maintain clarity in the figure. Some of the other differences are also statistically significant.}
	\label{fig:acc}
\end{figure}

The match-mismatch classification accuracy for each speech representation is shown in figure \ref{fig:acc}. Similar to the linear forward modeling results, we see that all the speech representations outperform the baseline VAD representation (p $<$ 0.01). This implies that the model is able to find neural tracking of this extra phonetic information in the EEG. Note that for all speech representations, training subject-specific models by fine-tuning increases the classification accuracy by around 5 \%. Since we have the same relative accuracy difference between representations in both the subject-independent and fine-tuned scenarios, we will only consider the fine-tuned model's results when comparing speech representations.

We observe that vowel-consonant onsets reach 84\% median accuracy compared to 82.5\% median accuracy of phone onsets (p $<$ 0.01). This supports the notion that discriminating between vowels and consonants helps the model extract useful information from the EEG to do the match-mismatch task. As with linear models, removing either vowel or consonant onsets decreases performance (p $<$ 0.01). Once again, we observe that vowel onsets are easier related to EEG than consonant onsets (p $<$ 0.01).
% This suggests that the model is able to, to some degree, extract distinguishing features between vowels and consonants.

The results further show that the more detailed phonetic representations, namely BPC and NPC onsets, do not surpass the performance of the vowel-consonant onsets representation. This suggests that the LSTM-based model is unable to extract this extra phonetic information from the EEG.

Another question we aimed to answer was whether vowel (syllable nucleus) onsets  are related better to EEG compared to syllable onsets. The results, as shown in figure \ref{fig:acc}, indicate that vowel onsets outperform syllable onsets (p $<$ 0.01).

Overall, the results are consistent with those of the forward modeling section. Vowel-consonant onsets outperforms phone onsets suggesting that the model is able to extract information about vowels and consonants from the EEG. Additionally,  vowel onsets were better related to EEG signals than syllable onsets. Furthermore, fine-tuning the model for each individual showed a 5\% improvement in accuracy.

\section{Discussion}

% Objectives of the study: 
%   general explanation of features used and rationale behind them

The purpose of this study was to investigate to which granularity level phonetic information of the presented stimulus can be related to the recorded EEG. To achieve this goal, we utilized various phonetic representations, including phone onsets, vowel-consonant onsets, syllable onsets, BPC onsets, and NPC onsets. We started from the most detailed phonetic representation, which was the NPC onsets, and progressively created representations with decreasing granularity of phonetic information. In other words, phone onsets contains the least information while NPC onsets has the most phonetic information. 
We included speech vs. silence information (VAD) to %as baseline information for each 
all representations, following the findings of \citet{monesi21_interspeech} that silences play a key role in the match-mismatch task. By doing so, we ensured that each speech representation contains speech vs. silence information, allowing us to determine that any improvements in accuracy or correlations are a result of the phonetic representation and not just VAD information present in the representation. 
We employed forward modeling (EEG prediction) and match-mismatch tasks to relate EEG recordings to speech representations.

% vowel vs. consonant distinctions can be related to EEG

We found that the vowel vs. consonant information is represented in the recorded EEG signals. Both the forward modeling and match-mismatch tasks indicated that including vowel vs. consonant information over phone onsets results in improved performance. This suggests that the difference between vowels and consonants, to some degree, is encoded in the recorded EEG signal. Previous research by \citet{khalighinejad_dynamic_2017} demonstrated that EEG responses of vowels differ from those of the consonant categories such as nasals, plosives, and fricatives at various time intervals relative to the phone onset. However, they compared the average time-locked EEG responses of each phonetic group, while our study utilized forward modeling and match-mismatch tasks to determine if phonetic representations can be related to EEG recordings.

% Previous research claims relating EEG to BPC
%   explaining that BPC contains onset information

Previous studies have reported that incorporating phonetic information, such as BPC (sometimes called phonetic features) on top of the spectrogram or envelope can lead to better EEG predictions \citep{di_liberto_low-frequency_2015,lesenfants_data-driven_2019} or match-mismatch accuracy \citep{monesi21_interspeech}, respectively. However, in these studies, a vowel vs. consonant onsets representation was not used. The key point is to use speech representations that contain the onset of each phone such that the duration of each phone is available, which was not the case in these studies. For example, in our prior work \citep{monesi21_interspeech}, we did use a vowel-consonant representation, but the speech representations, including vowel-consonant, were one-hot for the duration of the vowel or consonant. The problem with this sort of ``duration" based encoding, in contrast to ``onset" based encoding, is that sometimes the phone duration information is lost. For example, if there are two consecutive consonants, the model cannot know the duration of each consonant since it sees a one-hot vector for the whole duration of the two consonants. However, if one uses a phonetic representation such as BPC where consonants (and vowels) belong to different categories, the duration information will be available most of the time unless the consecutive consonants are from the same phonetic class. Therefore, it is not clear whether the reported neural tracking of BPC representation comes from relating actual distinction between phonetic classes to EEG or it comes from relating vowel-consonant onsets, which contains vowel vs. consonant distinction plus the duration of each phone, to EEG.

% Extra information of BPC and phoneme identity could not be related to EEG 
%    Effect of previously reported BPC is due to vowel-consonant onsets

To investigate the question, we made sure in our speech representations the duration of phones is available by using a temporal encoding based on phone onsets (i.e. one-hot vectors only at the onsets, see figure \ref{fig:speech_features}) as well as using speech representations such as vowel-consonant onsets and phone onsets. Our results in this study suggest that the actual contribution does not come from the distinction of broad phonetic classes but rather from the vowel vs. consonant onsets. More specifically, in both the forward modeling and match-mismatch tasks, vowel-consonant onsets performed equal or better than BPC and NPC representations indicating that the extra distinction between phonetic classes could not be related to EEG beyond vowel-consonant distinction. Furthermore, it seems that both the linear model in forward modeling and the LSTM-based model in the match-mismatch task struggle to work with more sparse speech representations such as NPC. One explanation might be that we need even more training data such that there are enough samples for some of the classes (phones) that occur rarely. This being said, this study already used a substantial EEG dataset and obtaining even a bigger dataset is quite challenging.

%syllable nucleus is better tracked than syllable onsets [peakRate paper ref]

We also compared syllable onsets to vowel (syllable nucleus) onsets. We found that in both forward modeling and match-mismatch tasks, vowel onsets were better related to the recorded EEG than syllable onsets. Our findings are consistent with those of \citet{oganian_speech_2019}, who used a speech representation based on a half-rectified speech envelope derivative (envelope peak rates) and reported that it outperformed the envelope in the forward modeling task. They showed that envelope peak rates are more aligned to vowel onsets than syllable onsets.

%- delay in vowel responses compared to consonant responses. correlation topoplots

In our analysis of the linear forward modeling task, we delved further into the TRF of speech representations. Our observations indicated that the majority of the representations displayed a positive peak before 100 ms. Our results were consistent with the TRFs of NPC (called phonemes) and BPC (phonetic features) reported by \citet{lesenfants_data-driven_2019} and \citet{di_liberto_low-frequency_2015}.
However, our baseline VAD representation was found to have two positive peaks at around 100 and 200 ms and a negative peak at 400 ms. The timing and polarity of the peaks differ from those reported in studies by \citet{di_liberto_low-frequency_2015} and \citet{lesenfants_data-driven_2019}, who found a negative peak at 80 ms and a positive peak at 150 ms for the speech envelope. The differences in the results of our study and the mentioned studies may stem from the difference in the speech representations used (VAD versus envelope) as well as the variation in the regularization matrices. We used an identity matrix in equation \ref{eqn:Rigde regression}, while the studies mentioned above used a matrix that penalizes the difference between neighboring terms \citep{crosse_multivariate_2016}. 
Additionally, we observed that the first peak of vowel onsets occurs after that of consonant onsets, which we hypothesize could be due to the fact that the energy of many consonants is concentrated closer to the onset, whereas vowel energy is more widely spread.

% summarize the main findings of the study

In summary, our study aimed to determine which granularity of phonetic information could be effectively related to EEG signals. We employed speech representations based on NPC onsets, preserving the duration between phones. Notably, our results revealed that the vowel-consonant onsets representation outperformed phone onsets in both forward modeling and match-mismatch tasks. This finding indicates that vowel vs. consonant information is represented in EEG. Next, we explored whether the extra phonetic information present in broad phonetic classes could be related to EEG. Interestingly, our results indicated that providing additional phonetic details beyond vowel vs. consonant did not yield improvements in the models' performance for either task. This suggests that the previously reported neural tracking associated with broad phonetic classes in studies such as \citep{lesenfants_data-driven_2019,di_liberto_low-frequency_2015,monesi21_interspeech} may not be attributed to the full phonetic information present in these phonetic classes, but rather to the information present in vowel vs. consonant onsets.

\section{Acknowledgement}
Funding for this project is provided by the KU Leuven Special Research Fund C24/18/099 (C2 project to Tom Francart and Hugo Van hamme) and the Flemish Government under ”Onderzoeksprogramma AI Vlaanderen”. This project has also received funding from the European Research Council (ERC) under the European Unions Horizon 2020 research and innovation programme (grant agreement No 637424, ERC starting Grant to Tom Francart).

% \section*{References}
% \begin{harvard}
%     \bibliographystyle{dcu}
%     \bibliography{refs.bib}
    
% \end{harvard}

\bibliographystyle{plainnat}
\newcommand{\newblock}{}
\bibliography{refs.bib}

\end{document}